\documentclass[twocolumn,showpacs,preprintnumbers,amsmath,amssymb,floatfix]{revtex4}

\usepackage{epsfig,psfrag}
\usepackage{color}
\usepackage{graphicx}
\usepackage{dcolumn}
\usepackage{bm}

\begin{document}

\title{Artifical atoms in interacting graphene quantum dots}

\author{Wolfgang H\"ausler$^{1,2}$ and Reinhold Egger$^1$}
\affiliation{
$^{1}$Institut f\"ur Theoretische Physik, Heinrich-Heine-Universit\"at, 
D-40225 D\"usseldorf, Germany \\
$^{2}$Physikalisches Institut, Albert-Ludwigs-Universit\"at, 
D-79104 Freiburg, Germany}

\date{\today}
\begin{abstract}
We describe the theory of few Coulomb-correlated electrons in a 
magnetic quantum dot formed in graphene.  While the 
corresponding nonrelativistic (Schr\"odinger) problem is well understood, 
a naive generalization to graphene's ``relativistic'' (Dirac-Weyl) spectrum
encounters divergencies and is ill-defined.  We employ Sucher's projection 
formalism to overcome these problems.  Exact diagonalization results
for the two-electron quantum dot, i.e., the artificial helium atom in 
graphene, are presented.
\end{abstract}


\maketitle

The recent spectacular progress in preparing and usefully employing
indivual carbon monolayers of graphene \cite{geim,castro} 
continues to stimulate much interest across different scientific
communities, including material science, applied physics, 
chemistry, condensed matter physics and mathematics.
Ballistic electronic motion with quantum coherence extending over micrometer 
distances has been achieved in several experiments, see, e.g., 
Ref.~\cite{andrei}.  The low-energy physics close to a single $K$ point 
can then be described by a two-component Dirac-Weyl Hamiltonian
 \cite{castro,foot},
\begin{equation}\label{diracweyl}
H_0 =v_{\rm F}\:\bm{\sigma}\cdot\left( \bm{p}-\frac{e}{c}\bm{A}\right)\;,
\end{equation}
suggesting an easily accessible condensed-matter 
realization of relativistic quantum mechanics. In Eq.~(\ref{diracweyl}),
$\bm{\sigma}$ denotes the vector of the first two Pauli
matrices for the ``isospin'' encoding the two sublattices, 
the Fermi velocity is $v_{\rm F}\approx 10^6$~m$/$sec,
and we include a static vector potential $\bm{A}(\bm{r})$
describing (possibly inhomogeneous) magnetic fields.
Since graphene's effective fine structure constant is $\alpha\approx 1$,
present interest has also turned to Coulomb interaction effects \cite{castro}.
According to recent Monte Carlo simulations \cite{mc} and analytical 
arguments \cite{khvesh}, sufficiently strong interactions may even open a 
sizeable bulk gap in the Dirac fermion spectrum.  

Here we study the properties of few Coulomb-correlated
electrons confined to a finite-size quantum dot formed in graphene.  
Using electrostatically formed quantum dots in semiconductor
devices,  such ``artificial atoms'' have been intensely  studied
over the past two decades, both experimentally 
\cite{kouwenhoven} and theoretically \cite{reimann}.
In graphene dots formed by electrostatic gating, however, carriers 
can escape due to the (recently observed \cite{stander}) 
Klein tunneling phenomenon, and at best quasi-bound 
states may appear \cite{efetov}.  An alternative is to employ 
lithographically defined quantum dots \cite{stampfer}, where
detailed information on ground- and excited-state properties has 
been obtained from transport spectroscopy.  Unfortunately, the 
boundary of lithographically fabricated graphene dots is 
rather disordered and difficult to model \cite{schnez2}.
On the other hand, suitable and realizable inhomogeneous
magnetic fields can confine Dirac fermions \cite{ademarti,haeusler1}, 
promising to yield tunable and well-defined magnetic graphene dots.
   
A more challenging difficulty to theory arises when trying to 
generalize Eq.~(\ref{diracweyl}) to a first-quantized many-particle description.
The first-quantized approach has turned out to be very efficient
and convenient for the case of Schr\"odinger electrons in 
semiconductor-based artificial atoms \cite{reimann}.
For the ``relativistic'' graphene case,
the problem arises from the unboundedness of Eq.~(\ref{diracweyl}), 
in contrast to the corresponding Schr\"odinger operator 
$(\bm{p}-\frac{e}{c}\bm{A})^2/2m^*$.  While Eq.~(\ref{diracweyl}) can 
still be used within effective single-particle approximations
such as the Hartree-Fock approach \cite{hf1},
variational schemes \cite{var}, or density-functional theory \cite{polini}, 
the full $N$-particle problem (for small $N>1$)
with Eq.~(\ref{diracweyl}) for the kinetic part suffers from the so-called
``Brown-Ravenhall disease'' \cite{brown,sucher}. Roughly speaking, the
unbounded spectrum allows particles to lose arbitrary amounts
of energy by transferring their energy in (real) scattering
events to other particles. The resulting divergent density of
states prohibits, for example, the direct use of exact
diagonalization methods. This difficulty of the
Dirac equation has been known for half a century \cite{brown}. To
``cure'' this ``disease'',  we follow a proposal by Sucher
\cite{sucher} and confine the Hilbert space to positive energy
eigenstates through suitably defined projectors, cf.\
Eq.~(\ref{lambdaplus}) below. While we formulate this approach
for the magnetic dot only, the general concepts remain applicable for almost
arbitrary graphene dots. The projection method then allows, for instance, to
apply numerical techniques to the relativistic $N$-particle problem.
In this work, we present exact diagonalization results for the many-body 
energy spectrum of the artificial helium atom ($N=2$) in graphene.

\begin{figure}[ht!]
\begin{minipage}[t]{8.0cm}
\epsfig{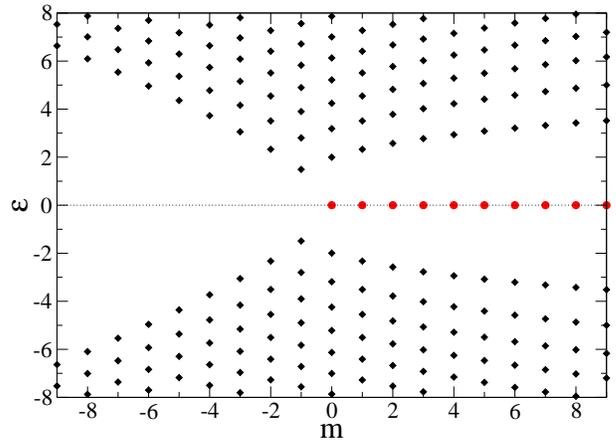}
\caption{\label{fig1} (Color online) Exact diagonalization 
results for the single-particle
eigenenergies $\varepsilon$ vs orbital angular momentum $m$
in a parabolic magnetic quantum dot in graphene [see Eq.~(\ref{spherical})].
The zero-energy levels are indicated as red filled circles, while the 
other levels are shown as black filled diamonds.  
}
\end{minipage}
\end{figure}

Let us first specify the model discussed here (we set $\hbar=1$). 
In cylinder coordinates, we consider the spherical and parabolic 
magnetic field profile oriented perpendicular to the graphene plane
(with ${\bm A}$ in symmetric gauge),
\begin{equation}
\bm{B}(r,\varphi) = \frac{c}{e} \omega_{\rm B}^2 r^2\bm{e}_z \; ,
\quad \bm{A}(r,\varphi)=\frac{1}{4}\omega^2_{\rm B} r^3
\left( \begin{array}{c}-\sin\varphi \\ \cos\varphi \\ 0 \end{array}\right ) .
\end{equation}
The inverse length scale $\sqrt{\omega_{\rm B}}$
tunes the field inhomogeneity.  The dimensionless radial coordinate is 
$\varrho=r\sqrt{\omega_{\rm B}}$, and energy ($\varepsilon$)
is measured in units of $v_{\rm F}\sqrt{\omega_{\rm B}}$. 
[Physical units are recovered from $\sqrt{B(\varrho)/{\rm Tesla}}= 
 ( v_{\rm F}\sqrt{\omega_{\rm B}}/ {\rm 26meV} ) \varrho.$]
Such magnetic profiles can be generated with reasonable accuracy
using suitable lithographically defined ferromagnetic films 
deposited on top of the graphene layer after formation of a protective oxide
layer \cite{heinzel}.  Upper and lower components of 
eigenspinors $|\psi_\nu^{(0)}\rangle$ to Eq.~(\ref{diracweyl}) must 
then differ by one orbital angular momentum quantum number ($m$) 
due to conserved total angular momentum \cite{ademarti}. 
With real functions $\phi_m(\varrho)$ and $\chi_{m+1}(\varrho)$,
the radial part of $|\psi_\nu^{(0)}\rangle$ is 
$\propto \left(\phi_m(\varrho), {\rm i} \ {\rm sgn}(\varepsilon)\ 
\chi_{m+1}(\varrho)\right)^T$, where Eq.~(\ref{diracweyl}) yields
the radial equations
\begin{equation}\label{spherical}
\left(\begin{matrix}
-\varepsilon & \partial_{\varrho}+\frac{m+1}{\varrho}-\frac{\varrho^3}{4}\cr
-\partial_{\varrho}+\frac{m}{\varrho}-\frac{\varrho^3}{4} & -\varepsilon
\end{matrix}\right) \left(\begin{matrix}
\phi_m(\varrho)\cr\chi_{m+1}(\varrho) \end{matrix}\right)=0\;,
\end{equation}
which cannot be solved analytically. We here carry out exact 
diagonalizations, later on including the Coulomb interaction,
and thus solve Eq.~(\ref{spherical}) numerically.
It is convenient to employ the  Darwin-Fock states (with integer
$n\geq 0$ and the Laguerre polynomials $L_{n}^{|m|}$) \cite{reimann}
\begin{equation}\label{darwinfock}
\Phi_{nm}^{(\lambda)}(\varrho,\varphi)=
\frac{{\rm e}^{{\rm i}m\varphi}}{\sqrt{2\pi}}
\sqrt{\frac{2\lambda n!}{(n+|m|)!}}{\rm e}^{-\lambda\varrho^2/2}
(\sqrt{\lambda}\varrho)^{|m|}L_n^{|m|}(\lambda\varrho^2)
\end{equation}
as complete orthonormal function set in two dimensions (2D) to
expand $\phi_m$ and $\chi_{m+1}$ in Eq.~(\ref{spherical}). In
Eq.~(\ref{darwinfock}), we have included an additional tunable
width parameter $\lambda$, which can be optimized \cite{foot2} to 
reduce the number of required basis functions $0\le n\le n_{\rm max}$
when approximating the $|\psi_\nu^{(0)}\rangle$ to the desired accuracy. 
Figure \ref{fig1} displays the resulting eigenenergies 
as a function of the orbital angular momentum $m$.
As expected, the spectrum is electron-hole symmetric, and
for $m\ge 0$, a zero-energy level develops. This zero-energy level
is non-dispersing (precisely like a quantum Hall level), despite
of the inhomogeneous magnetic field which implies
the nontrivial $m$-dependence of all other energy levels.
Note that for the corresponding Schr\"odinger case with a 
parabolic magnetic field, the zero-energy level is absent.

Next we consider $N$ interacting electrons in such a graphene dot. 
A naive approach is to consider the first-quantized Hamiltonian 
\begin{equation}\label{naive}
H = v_{\rm F}\sum_{j=1}^N\bm{\sigma}_j\cdot[\bm{p}_j-\bm{A}(\bm{r}_j)]+
\sum_{i<j} \frac{\alpha v_{\rm F}}{|\bm{r}_i-\bm{r}_j|}\;,
\end{equation}
where the fine structure constant is $\alpha=e^2/(\kappa_0 v_{\rm F})$.
For typical substrate materials, the dielectric constant is 
$\kappa_0\approx 1.4$ to $4.7$, resulting in $\alpha\approx 0.6$ to 2.
We mention in passing that the ``Wigner molecule'' regime \cite{egger} 
seems out of reach in graphene dots, since both the kinetic and
the potential energy show identical scaling when changing the
density \cite{noWC}.  Moreover, we neglect the Zeeman term which 
is very small in graphene \cite{haeusler1}.  Up to the spin and $K$-point
indices \cite{footnote2}, many-body spinors then have $2^N$ components.
For the related Schr\"odinger dot ($H_{\rm S}$), confinement of electrons 
is usually achieved by a parabolic electrostatic 
potential \cite{kouwenhoven}, and 
the many-particle description analogous to Eq.~(\ref{naive}) simplifies
considerably owing to the generalized Kohn theorem \cite{chaplik}.
According to this theorem, $H_{\rm S}=H_{\rm cm}+H_{\rm
rel}$ separates into two commuting parts describing center-of-mass
($H_{\rm cm}$) and relative ($H_{\rm rel})$ motion. 
Then $H_{\rm cm}$ is just a 2D harmonic oscillator, 
while $H_{\rm rel}$ contains all Coulomb interaction effects. 
In addition, $H_{\rm rel}$ conserves angular momentum, as does $H_{\rm cm}$. 
Taking $N=2$ as example, in effect only a 1D quantum problem for the 
radial motion of $H_{\rm rel}$ remains to be solved.
In contrast, Eq.~(\ref{naive}) does not benefit from Kohn's
theorem, and only the total angular momentum remains conserved as
dictated by isotropy. Therefore, while the additional spinor
structure already increases the rank of the Hamiltonian matrix 
in the Dirac-Weyl case by a factor $2^N$, 
the rank grows even more severely because neither $H_{\rm cm}$ nor
a conserved angular momentum of $H_{\rm rel}$ can be separated
off the problem.  For $N=2$ (graphene helium), we needed to include states
up to $n_{\rm max}\approx 14$ to reach sufficient accuracy.
In addition, contrary to the Schr\"odinger problem, particles
may now exchange relative angular momentum $\Delta m$ through the interaction.
Owing to the exponential decay of Coulomb matrix elements with $|\Delta m|$,
it is sufficient to take $|\Delta m|\leq 3$, yielding an
additional factor $7^{N-1}$ to the matrix size (for $N$ particles).
For $N=2$, we then need to include $14^2\times 7\times 2^2=5488$
product basis states in total.

Let us then address the more fundamental difficulty arising already for $N=2$
when naively using Eq.~(\ref{naive}).  A closely related problem 
has been pointed out by Brown and Ravenhall \cite{brown} in a 
relativistic treatment of the helium atom: the Dirac equation analogous to 
Eq.~(\ref{naive}) does not possess normalizable antisymmetric eigenstates 
in the two-particle Hilbert space.  This failure has its origin in the
unbounded spectrum of the Dirac Hamiltonian, which allows for
unlimited energy exchange among the particles. As a result, 
the density of two-particle states increases with Hilbert space
dimension and ultimately diverges.  This causes, e.g., 
divergent contributions in second-order perturbation theory. 
In consequence, neither the true Dirac equation nor the two-component 
variant (\ref{diracweyl}) for graphene allow for naive 
many-particle generalizations such 
as Eq.~(\ref{naive}). To overcome this deficiency, Sucher \cite{sucher} 
proposed to restrict the (anti-symmetrized) product
Hilbert space to the positive energy eigenspace for each
particle by means of a suitable projector $\Lambda_+$.
We thus consider a situation with chemical potential $\mu=0^+$ just above
zero, where all non-positive energy (hole) states up to (and including) 
zero energy are filled.  As long as a finite energy gap 
separates the relevant positive energies (which can then be occupied
by the $N$ electrons under consideration) from the filled
non-positive energy levels, such a projection seems physically
sensible, at least for interaction strengths not exceeding this gap. 
In natural helium, the mass gap ensures the validity of such an approach.
On the other hand, in a homogeneous 2D graphene sheet,  weakly interacting
fermions are gapless and Sucher's approach does not apply.
For our finite-size quantum dot, however, there is a 
finite gap between $\varepsilon=0$ and the lowest positive-energy level, 
see Fig.~\ref{fig1}.  
We therefore consider two \textit{additional} electrons confined in the 
quantum dot, residing above a filled Dirac sea 
with $\mu=0^+$. The projector proposed by Sucher is expressed as
\begin{equation}\label{lambdaplus}
\Lambda_+=\Lambda_+^{(1)}\otimes\Lambda_+^{(2)} \ , \quad
\Lambda_+^{(j)}=\sum_{\nu\in I}|\psi_{\nu}^{(j,0)}\rangle
\langle\psi_{\nu}^{(j,0)}| \ ,
\end{equation}
where the sum is restricted to strictly positive single-particle 
energies $E_{\nu\in I}^{(j,0)}>0$ (for particle $j=1,2$) indexed by $\nu\in I$, 
see Fig.~\ref{fig1}, with 
corresponding eigenspinors $|\psi_{\nu}^{(j,0)}\rangle$.
With $H$ in Eq.~(\ref{naive}), the projected Hamiltonian 
$H_{\rm D} = \Lambda_+ H \Lambda_+ $ is well-behaved and exhibits a finite 
density of two-particle states which does not increase with Hilbert 
 space size. Throughout the experimentally relevant regime,
 $\alpha\alt 2$, interactions are not strong enough to induce a
breakdown of this projection approach.  

\begin{figure}[ht!]
\begin{minipage}[t]{8.0cm}
\epsfig{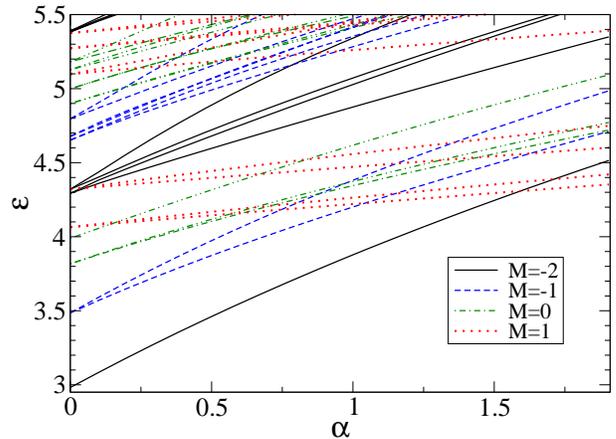}
\caption{\label{fig2} (Color online) Exact diagonalization results
for the energy spectrum of graphene artificial helium
 vs fine structure constant $\alpha$ for different total
angular momenta $M$.  States with $M=-2$ are shown as
black solid, $M=-1$ as blue dashed, $M=0$ as green dot-dashed,
and $M=1$ as red dotted curves.  }
\end{minipage}
\end{figure}

We have carried out exact diagonalizations of $H_{\rm D}$
for $N=2$ using a two-particle product basis of Darwin-Fock states 
(\ref{darwinfock}).  This basis has the advantage of allowing
to analytically express the matrix elements of the two-particle interaction 
operator in Eq.~(\ref{naive}) in terms of finite sums and products, i.e., no
quadratures nor truncations of infinite sums are necessary; for
the corresponding (lengthy) expressions, see Ref.~\cite{halonen}.
Their numerical evaluation involves taking small differences of 
huge numbers, the latter increasing as $n_{\rm max} !$ with the
number of kept Darwin-Fock states. We employed
algorithms allowing for number manipulations of arbitrary
precision and used 30-digits accuracy.
The resulting energy spectrum of artificial helium in a magnetic 
graphene quantum dot is shown in Fig.~\ref{fig2} for conserved total 
orbital angular momenta $M=m_1+m_2=-2,-1,0,1$.  These values include the 
ground state for $\alpha<2$.  All levels rise when increasing the 
(repulsive) interaction strength.  This also holds true for the 
hole states (not displayed in Fig.~\ref{fig2}), which, however, 
cross $\varepsilon=0$ only for $\alpha\gg 2$.  
For $\alpha<2$, interaction  matrix elements
indeed remain much smaller than the energy difference 
$\Delta\varepsilon\simeq 2.98165$ between the lowest two-particle state 
(for $M=-2$ at $\alpha=0$) and the zero-energy level, 
\textit{a posteriori} justifying Sucher's approach here.  
Figure \ref{fig2} reveals that states with larger
total angular momentum $M$, or higher excited states, tend to 
increase less in energy with $\alpha$ as
compared to the $M=-2$ ground state. This is a consequence of
the larger spatial extent of excited-state wavefunctions, with a
reduced Coulomb repulsion between the electrons. Particularly
striking is the shallow increase of the lowest $M=1$ energy level,
which even becomes lower in energy than the $M=-2$ level for
$\alpha\agt 1.6$.
Approximating this level at $\alpha=0$ by Darwin-Fock levels,
one of the two particles is seen to have $m=2$ for the lower spinor component,
cf.~Eq.~(\ref{spherical}), causing a significantly larger spread of this 
part of the wavefunction compared to $M=-2,-1,0$.  
Figure \ref{fig2} also reveals nontrivial \textit{spin} physics.
In the presence of interactions $(\alpha > 0)$, 
doubly degenerate noninteracting ($\alpha=0$) energy levels 
will split into a spin-triplet ($S=1$) state of lower energy 
and a spin-singlet ($S=0$) state of higher energy, 
in accordance with Hund's rule. The triplet states are approximately
(see below) Zeeman-degenerate.  
Singly degenerate $\alpha=0$ levels, such as the $M=-2$ ground
state (for small $\alpha$), are $S=0$ states and remain unsplit 
for $\alpha>0$. Thus we expect singlet-triplet ground-state spin 
transitions to occur within $0<\alpha <2$, as the one seen in
Fig.~\ref{fig2} at $\alpha\approx 1.6$.

Finally, we remark on \textit{optical transitions} between the many-body energy
levels in Fig.~\ref{fig2}. For the electrostatically defined parabolic
Schr\"odinger quantum dot, the generalized Kohn theorem implies that 
Coulomb interactions can never affect optical transitions because the 
dipole operator $\:\sum_{j=1}^N\bm{r}_j\:$ acts exclusively on the
eigenspace of $H_{\rm cm}$. Therefore optical spectra
just reflect the harmonic excitations of the center-of-mass motion 
\cite{reimann}.  However, in our magnetic graphene dot,
Kohn's theorem is ineffective and optical transitions between different
many-body levels in Fig.~\ref{fig2} are possible, thereby allowing to 
optically probe interaction physics.  Note that magnetic fields are usually 
assumed homogeneous such that photons cannot change the total spin $S$ of
the charged many-particle system in electrical dipole transitions.
While this would prohibit all transitions between states with different $S$,
the \textit{inhomogeneous} magnetic field here (slightly) mixes the 
$S_z=0$ components of $S=0$ and $S=1$ levels. We estimate the
amount of this mixing by the variation of the Zeeman energy across 
the spatial extent of the wavefunction 
compared to the level separations of $H_{\rm D}$. As a first
estimate, compare the Zeeman energy $\Delta_Z$ at the
maximum of the charge-density distribution with the orbital
(Landau) energy $\Delta_L$ at this point,
$\Delta_Z/\Delta_L= g\mu_{\rm B}B/\sqrt{2ecB}
\simeq 10^{-5}\sqrt{B/{\rm Tesla}}$.
While this is small, the Zeeman energy
variations can easily exceed orbital level separations  near spin
singlet-triplet degeneracies, e.g., for $\alpha\ll 1$ or close to 
level crossings in Fig.~\ref{fig2}, resulting
in a strong spin mixing. The corresponding transitions are then 
optically allowed.

To conclude, we have presented the theory of few interacting electrons
in a (magnetically confined) graphene quantum dot.  The low-energy 
Dirac-Weyl spectrum of graphene suggests that one can realize
relativistic artificial atoms in this setting. While a naive 
formulation encounters conceptual difficulties related to the 
unboundedness of the Dirac-Weyl Hamiltonian, by virtue of Sucher's
projection operator approach, a consistent and accurate theory can 
be given.  We have presented exact diagonalization results for the 
energy spectra of artificial helium, where we predict
singlet-triplet ground-state spin transitions to occur for $\alpha<2$.
Moreover, the  reported many-body levels can be experimentally probed by 
optical spectroscopy.--- We thank H. Siedentop for drawing our attention to Refs.~\cite{brown}
and \cite{sucher}, and acknowledge discussions with A.  De Martino, 
K. Ensslin, and T. Heinzel. 
This work was supported by the DFG (SFB TR/12) and by the ESF network INSTANS.


\begin{thebibliography}{99}

\bibitem{geim}
A.K. Geim and K.S. Novoselov, Nat. Mater. {\bf 6}, 183 (2007).

\bibitem{castro}
A.H. Castro Neto, F. Guinea, N.M.R. Peres, K.S. Novoselov, and A. Geim,
Rev. Mod. Phys. {\bf 81}, 109 (2009).

\bibitem{andrei}
X. Du, I. Skachko, A. Barker, and E.Y. Andrei, Nat. Nanotech. {\bf 3},
491 (2008); K.I. Bolotin, K.J. Sikes, J. Hone, H.L. Stormer,
and P. Kim, Phys. Rev. Lett. {\bf 101}, 096802 (2008).

\bibitem{foot}
The true electronic spin is kept implicit, and there is a similar
Hamiltonian for the other $K$ point.

\bibitem{mc}
J.E. Drut and T.A. L\"ahde, Phys. Rev. Lett. {\bf 102}, 026802 (2009);
Phys. Rev. B {\bf 79}, 165425 (2009).

\bibitem{khvesh}
D.V. Khveshchenko, J. Phys. Cond. Matt. {\bf 21}, 075303 (2009).

\bibitem{kouwenhoven}
L.P. Kouwenhoven, D.G. Austing, and S. Tarucha,
Rep. Prog. Phys. {\bf 64}, 701 (2001).

\bibitem{reimann}
S.M. Reimann and M. Manninen, Rev. Mod. Phys. {\bf 74}, 1283 (2002). 

\bibitem{stander}
N. Standar, B. Huard, and D. Goldhaber-Gordon, Phys. Rev. Lett.
{\bf 102}, 026807 (2009); 
A.F. Young and P. Kim, Nat. Phys. {\bf 5}, 222 (2009).

\bibitem{efetov}
J. Milton Pereira, Jr., V. Mlinar, F.M. Peeters, and P. Vasilopoulos,
Phys. Rev. B {\bf 74}, 045424 (2006);
P.G. Silvestrov and K.B. Efetov, Phys. Rev. Lett. {\bf 98}, 016802 (2007);
H.-Y. Chen, V. Apalkov, and T. Chakraborty, \textit{ibid.} 
{\bf 98}, 186803 (2007). 

\bibitem{stampfer}
C. Stampfer, J. G\"uttinger, F. Molitor, D. Graf, T. Ihn, and K. Ensslin,
Appl. Phys. Lett. {\bf 92}, 012102 (2008);
L.A. Ponomarenko \textit{et al.}, Science {\bf 320}, 356 (2008);
S. Schnez \textit{et al.}, Appl. Phys. Lett. {\bf 94}, 012107 (2009);
J. G\"uttinger \textit{et al.}, preprint arXiv:0904.3506.

\bibitem{schnez2}
S. Schnez, K. Ensslin, M. Sigrist, and T. Ihn, Phys. Rev. B {\bf 78},
195427 (2008); J. Wurm, A. Rycerz, I. Adagideli, M. Wimmer, K. Richter, 
and H.U. Baranger, Phys. Rev. Lett. {\bf 102}, 056806 (2009). 

\bibitem{ademarti}
A. De Martino, L. Dell'Anna, and R. Egger,  Phys. Rev. Lett. 
{\bf 98}, 066802 (2007); Sol. St. Comm. {\bf 144}, 547 (2007).

\bibitem{haeusler1}
W. H\"ausler, A. De Martino, T.K. Ghosh, and R. Egger,
Phys. Rev. B {\bf 78}, 165402 (2008).

\bibitem{hf1}
B. Wunsch, T. Stauber, and F. Guinea, Phys. Rev. B {\bf 77}, 035316 (2008);
M. Ezawa, \textit{ibid.} {\bf 77}, 155411 (2008).

\bibitem{var}
I. Romanovsky, C. Yannouleas, and U. Landman, Phys. Rev. B {\bf 79}, 
075311 (2009).

\bibitem{polini}
M. Polini, A. Tomadin, R. Asgari, and A.H. MacDonald,
Phys. Rev. B {\bf 78}, 115426 (2008). 

\bibitem{brown}
G.E. Brown and D.G. Ravenhall, Proc. R. Soc. London Ser. A
{\bf 208}, 552 (1951).

\bibitem{sucher}
J. Sucher, Phys. Rev. {\bf 107}, 1448 (1957);
Phys. Rev. {\bf 109}, 1010 (1958);
Phys. Rev. A {\bf 22}, 348 (1980);
Int. J. Quantum Chem. {\bf 25}, 3 (1984).

\bibitem{heinzel}
T. Heinzel, private communication.

\bibitem{foot2} In principle, each eigenstate $|\psi_{\nu}^{(0)}\rangle$ can be
individually optimized for $\lambda=\lambda(\nu)$, which can be expressed 
analytically as a fourth-order root. However, choosing $\lambda=1.7$ for
all levels is sufficient for the results presented here.

\bibitem{egger}
R. Egger, W. H\"ausler, C.H. Mak, and H. Grabert, Phys. Rev. Lett.
{\bf 82}, 3320 (1999);  {\bf 83}, 462(E) (1999).

\bibitem{noWC}
H.P. Dahal, Y.N. Joglekar, K.S. Bedell, and A.V. Balatsky,
Phys. Rev. B {\bf 74}, 233405 (2006).

\bibitem{footnote2}
$K$-$K'$ scattering can be neglected for the long-range 
Coulomb interaction. Assuming that the interaction
does not discriminate among sublattices at long wavelengths,
the last term in Eq.~(\ref{naive}) is diagonal in isospin
space and we can map wave functions between the $K$ points.
The spectrum is then independent of how particles 
are distributed over the two $K$ points.

\bibitem{chaplik}
A.O. Govorov and A.V. Chaplik, JETP Lett. {\bf 52}, 31 (1990).

\bibitem{halonen}
V. Halonen, T. Chakraborty, and P. Pietil\"ainen, Phys. Rev.  B {\bf 45},
 5980 (1992).


\end{thebibliography}
\end{document}